\documentclass[rnote]{aa} 
\usepackage{graphicx}
\usepackage{natbib}

\usepackage{txfonts}
\begin{document}
\title{Near-IR integral-field spectroscopy of 
the companion to GQ Lup \thanks{Based on observations collected at the European
Southern Observatory, Chile, in programmes 275.C-5033(A) and 077.C-0264(A).}}

\author{
Andreas Seifahrt \inst{1,2} \and
Ralph Neuh\"auser \inst{2} \and
Peter H. Hauschildt \inst{3}}
          
   \institute{
European Southern Observatory, Karl-Schwarzschild-Str. 2, 85748 Garching, Germany \and
Astrophysikalisches Institut, Universit\"at Jena, Schillerg\"asschen 2-3, 07745 Jena, Germany \and
Hamburger Sternwarte, Gojenbergsweg 112, 21029 Hamburg, Germany}

  \offprints{Andreas Seifahrt, \email{aseifahr@eso.org}}

   \date{Received 28 September 2006; Accepted 07 December 2006}
 
  \abstract
{The first substellar companion of possibly planetary mass around a normal star, 
GQ Lup, has been directly imaged (Neuh\"auser et al., 2005). Besides the unknown formation 
history, the mass of such an object is a criterion to decide about its true nature.}
{We aim to determine the physical properties of 
the GQ Lup companion
- effective temperate ($T_\mathrm{eff}$) and surface gravity ($\log{g}$), 
and thus its mass independently from evolutionary models.
}
{We use the adaptive optics integral-field spectrograph SINFONI at the VLT for near-infrared 
spectroscopy from 1.1 to 2.5 $\mathrm{\mu m}$ with a resolution of R = 2500--4000. 
We compare these spectra with synthetic atmospheric models (GAIA v2.0 cond).}
{From the complete set of spectra we find a consistent effective temperature and surface gravity of
$T_\mathrm{eff} = 2650 \pm 100~\mathrm{K}$ and $\log{g} = 3.7 \pm 0.5~\mathrm{dex}$. 
Combined with a slightly revised luminosity of $\log{L/L_{\sun}} = -2.25 \pm 0.24$~dex 
for the companion, we determine a radius of $R = 3.50^{+1.50}_{-1.03}~\mathrm{R_{Jup}}$ 
and thus a mass of $\sim 25 \mathrm{M_{Jup}}$. The uncertainty of this value is rather high.
Due to the large uncertainty of the surface gravity, the mass could range from 4 to 155 $\mathrm{M_{Jup}}$.
By comparing the paramaters of the companion of GQ Lup to the ones of \object{2MASS J05352184-0546085}, published by Stassun et al. (2006), we conclude that the companion to GQ Lup A has a mass lower than 
$36 \pm 3$ $\mathrm{M_{Jup}}$.}
{}

   \maketitle
%

\section{Introduction}
\citet{neuh05} and \citet{Mugi} presented a faint common proper motion companion to 
the $1\pm1$ Myr old T-Tauri star \object{GQ Lup}. Based on both, 
a low resolution NACO spectrum (R$\sim$700) in comparison to early GAIA model
atmospheres \citep{peter05} and evolutionary models \citep{wuchterl}, the companion was 
found to have spectral type M9 to L4 and a mass of up to a few Jupiter masses.

The mass estimate was, however, model-dependent. Due to the young age of GQ Lup, the most 
widely used evolutionary models from the Tucson group \citep{burrows93} and the Lyon group 
\citep{dusty} for low mass objects are not applicable, see, e.g., \citet{tooyoung}.

The direct determination of the mass from effective temperature, surface gravity and luminosity
was hampered by the low signal-to-noise ratio and low resolution of the NACO K-band spectrum that 
yielded $T_\mathrm{eff}$ and $\log{g}$ with wide error margins only. Moreover, the fit of 
the observed spectrum against a grid of synthetic spectra from GAIA dusty atmosphere models 
did not deliver consistent results for the CO band features and the overall slope of the spectrum 
(see e.g. \citet{LaPalma}).

It was thus desirable to obtain new near-infrared spectra with higher signal-to-noise ratio and
higher resolution over a wider spectral range. Because of the large magnitude difference 
between GQ Lup A and its companion and their small separation of $\sim$0.7\arcsec the use of 
adaptive optics systems was mandatory. Hence, we used the adaptive optics integral-field 
spectrograph SINFONI at the Very Large Telescope (VLT) that provides a spectral resolution of 
R$\sim$2500 in the J-band and R$\sim$4000 in the H and K-band. 

We present the observations and describe the data reduction in section 2. The spectral synthesis
is outlined in section 3. We present the results of the fitting process in section 4, where we end
with conclusions and a short discussion about the nature of the GQ Lup companion.

\section{Observations and data reduction}

\subsection{Observations}

SINFONI (Spectrograph for INfrared Field Observations) is a combination of two instruments.
First, a MACAO (Multi-Application Curvature Adaptive Optics) type curvature adaptive optics 
(AO) module with visual wavefront sensor and 60 actuators. Second, a mid-resolution near-infrared 
spectrograph with an integral field unit: SPIFFI 
(SPectrograph for Infrared Faint Field Imaging), see \citet{sinfoniI} and \citet{sinfoniII}.
The advantage of this instrument that combines the spatial resolving power of an 8m telescope 
with a mid-resolution spectrograph is the absence of wavelength dependent slit losses that 
occur on normal spectrographs with a narrow entrance slit, especially when combined with AO. 
 
The first observations of the GQ Lup companion have been carried out in K band in the night of Sept 16, 2005. 
The observations followed the standard scheme of sky nodding for background subtraction. 
Eight nodding cycles with a integration time of 300s per frame where obtained. We have chosen the 
smallest pixel scale (12.5 $\times$ 25.0 mas) of the instrument, yielding a field of view of 
0.8\arcsec $\times$ 0.8\arcsec\ to optimally sample the diffraction limited core of the target PSF. 
The bright primary, GQ Lup A, was used as the AO guide star and was placed outside the FOV. 
The companion was centered in the field. 

The DIMM\footnote{Seeing value in the optical ($\lambda\sim$500nm), measured at zenith.}
seeing during the science observations was 0.8--1.1\arcsec. The strehl ratio, computed on 
a short integration of GQ Lup A was $\sim$40\% and more than 99\% of the energy was 
encircled in a core of $\sim$200~mas FWHM. 

H and J band observations have been obtained in the nights of April 24 and September 18, 2006, 
respectively. Ten (nine) target-sky nodding cycles of 300s exposure time have been taken in the 
H and J band, respectively. Five frames in each band had a sufficiently high signal-to-noise. 

The DIMM seeing for the five useful H band observations was 1.0--1.4\arcsec. The strehl ratio, 
measured similarly as for the K band, was $\sim$20\% and more than 99\% of the energy was 
encircled in a core of $\sim$250~mas FWHM.

No PSF calibration was possible for the J band since no observation of GQ Lup A was taken. 
In situ strehl computations on the companion are not reliable because of the background contamination from GQ Lup A. 
We therefore judge from the seeing during the J band observations (0.8--1.1\arcsec), 
the airmass and the performance of MACAO on a strehl of 5--10\%. The core size of the 99\% quartile 
of the encircled energy is estimated to be of the order of 350~mas FWHM. 

\subsection {Data reduction}

The data format of SINFONI is highly complex. The FOV of the instrument is sliced into 32
slitlets and remapped onto a pseudo-slit and dispersed by a grating. The original FOV has to be 
reconstructed from this spectrum. The success of this process depends on the proper identification 
of the slitlet positions on the detector and the correction of nonlinear distortions. 

The first step of the data reduction was thus performed by using the SINFONI data reduction pipeline 
version 1.3 offered by ESO \citep{yves}. The reduction routines of this pipeline where developed 
by the SINFONI consortium \citep{SPRED}. In this first step the images are cleaned from bad pixels, 
flat fielded and wavelength calibrated. A distortion correction is applied and the the FOV of SINFONI 
is reconstructed. The output of this procedure is a 3D fits cube containing about 2000 images of 
the source in wavelength steps according to the chosen setting of the grating.

\begin{figure}[t!]
   \resizebox{\hsize}{!}{
\includegraphics[bb=80 170 532 622,angle=90,clip]{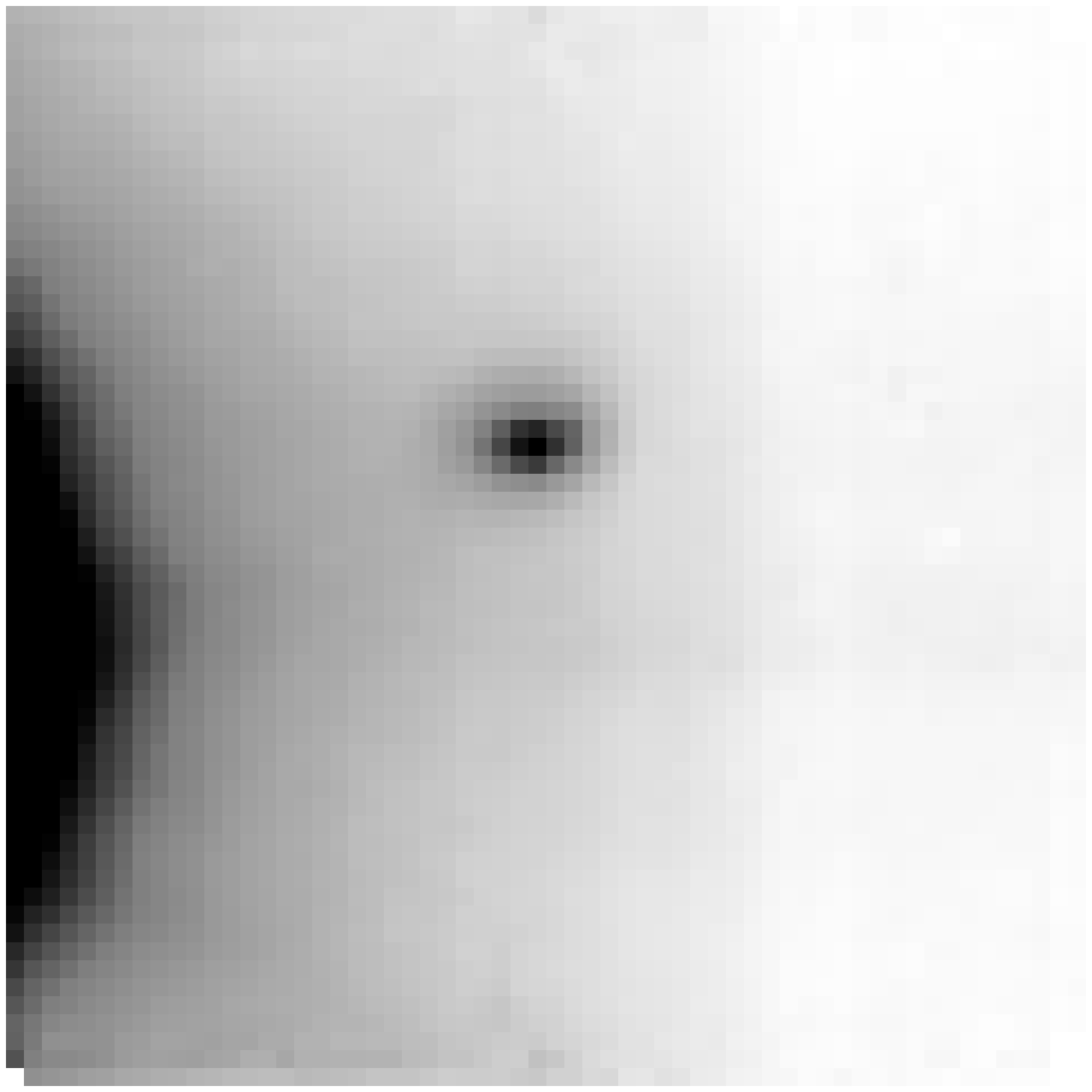}
\includegraphics[bb=80 170 532 622,angle=90,clip]{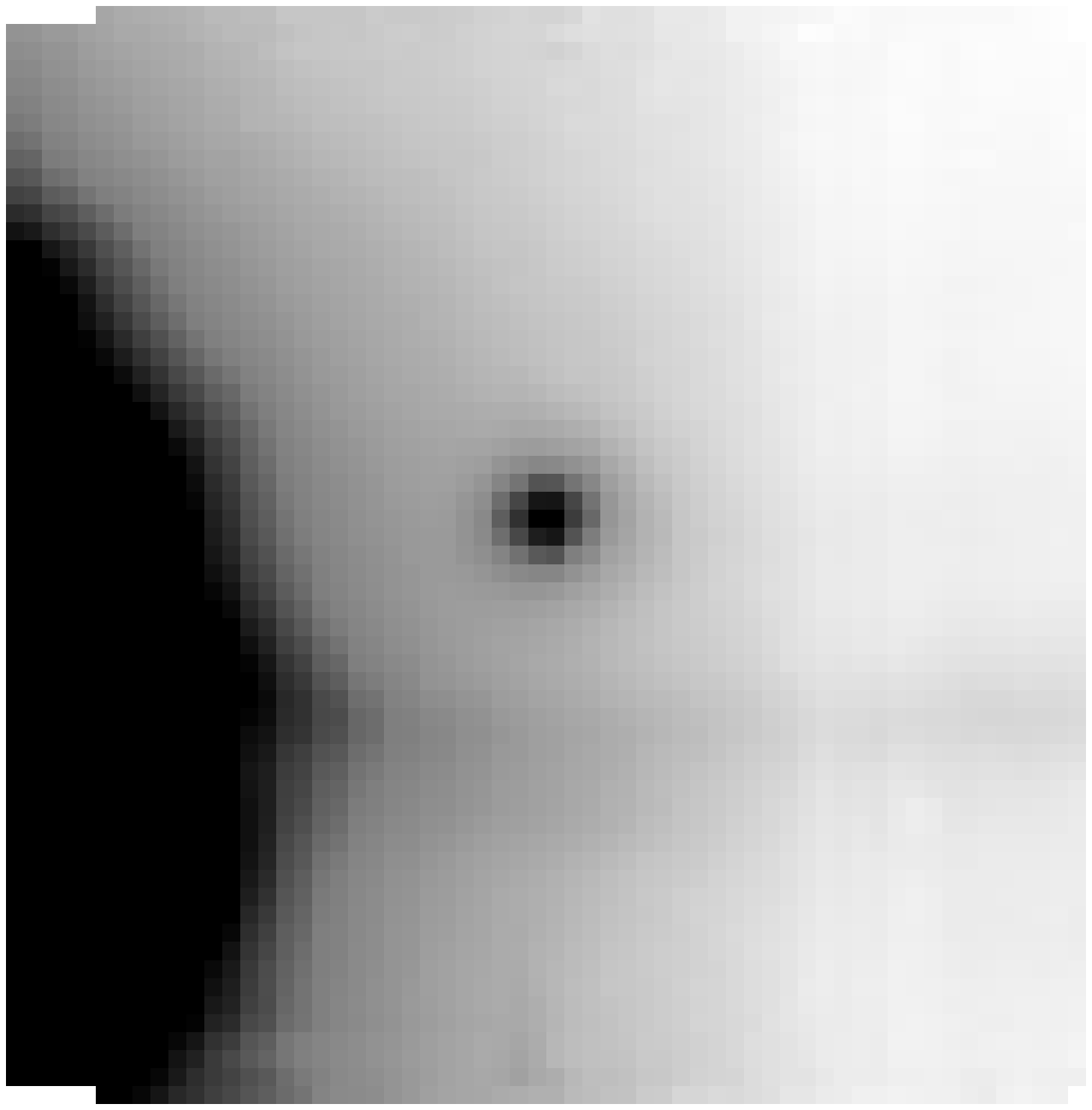}
\includegraphics[bb=80 170 532 622,angle=90,clip]{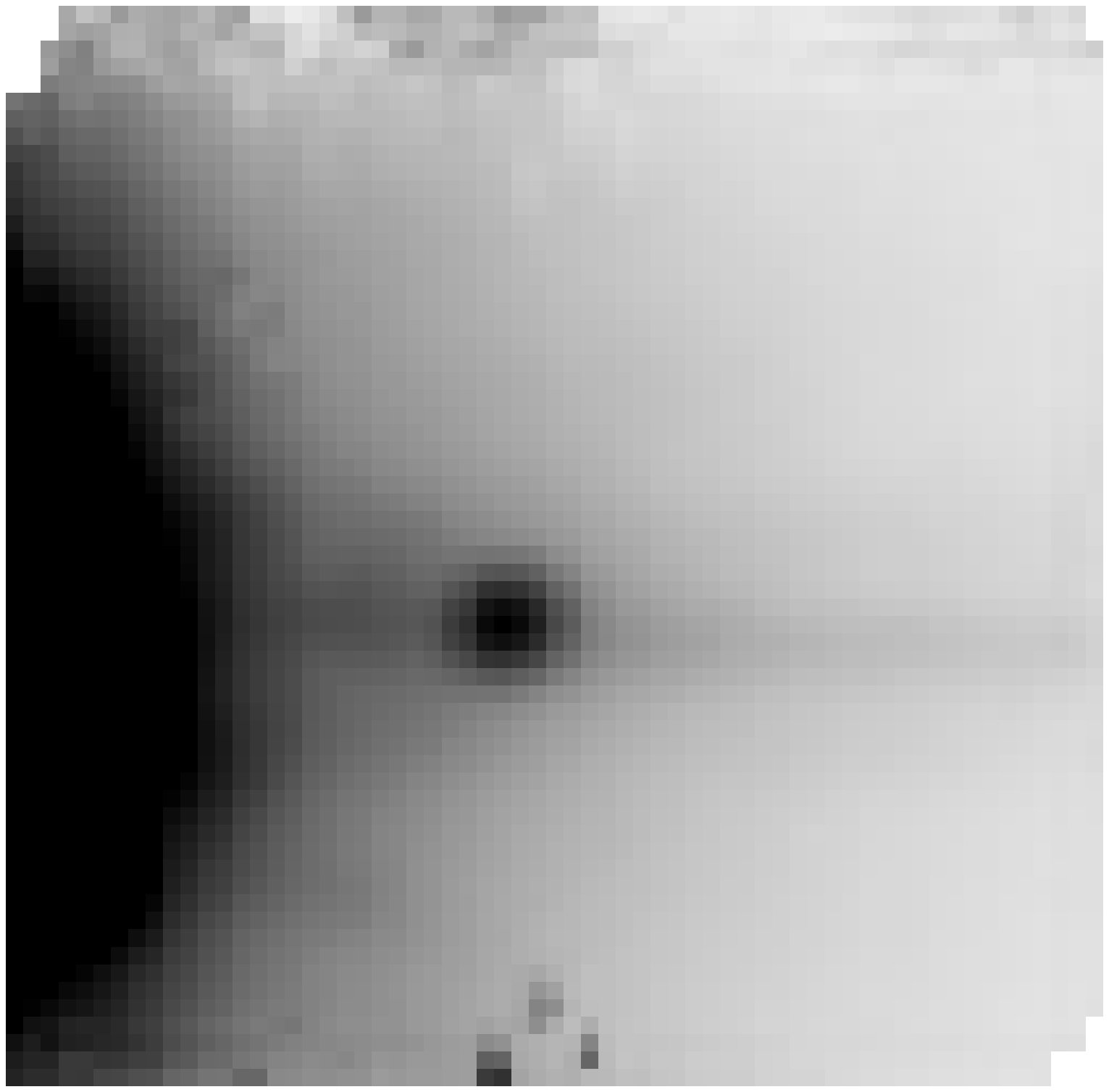}
}
   \resizebox{\hsize}{!}{
\includegraphics[bb=80 170 532 622,angle=90,clip]{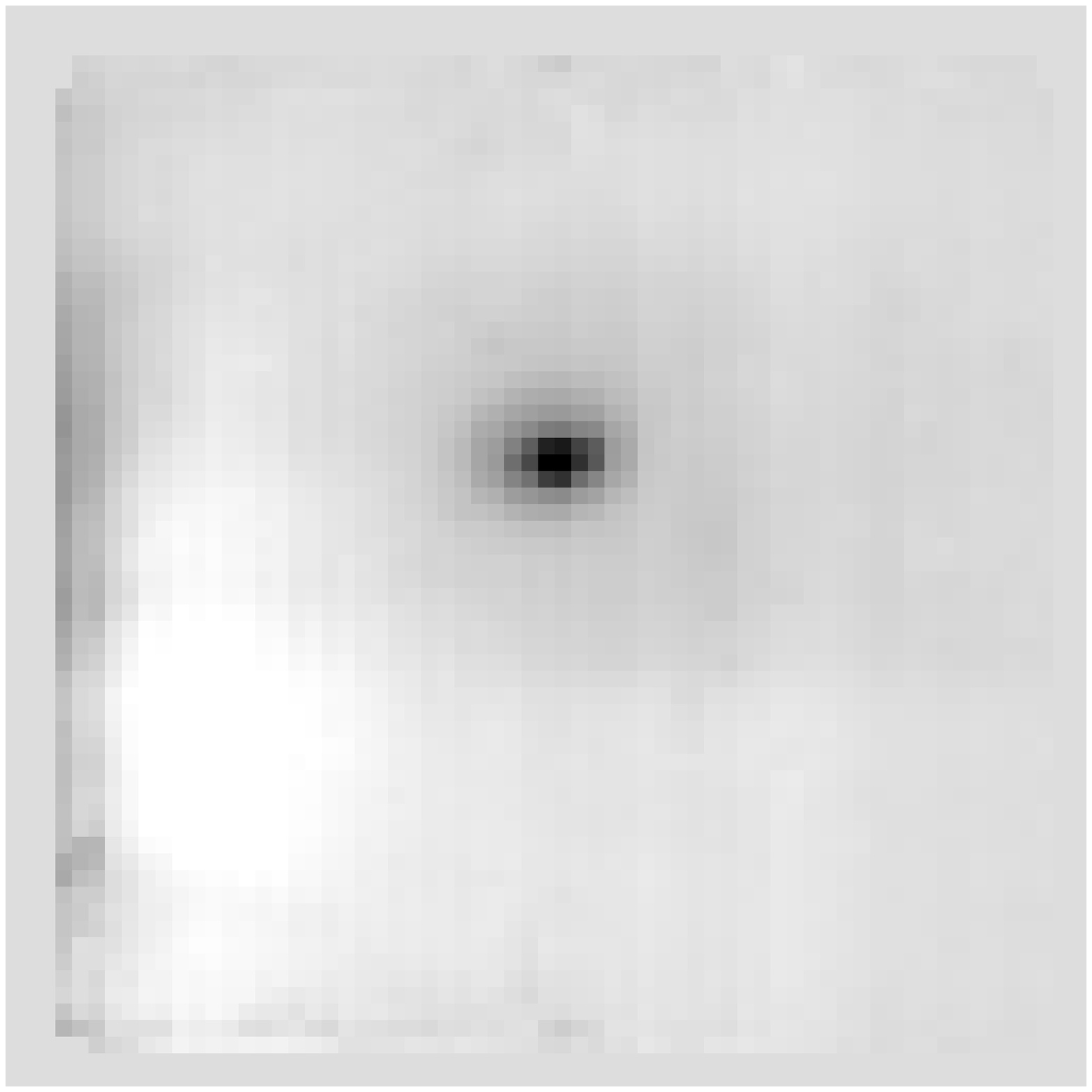}
\includegraphics[bb=80 170 532 622,angle=90,clip]{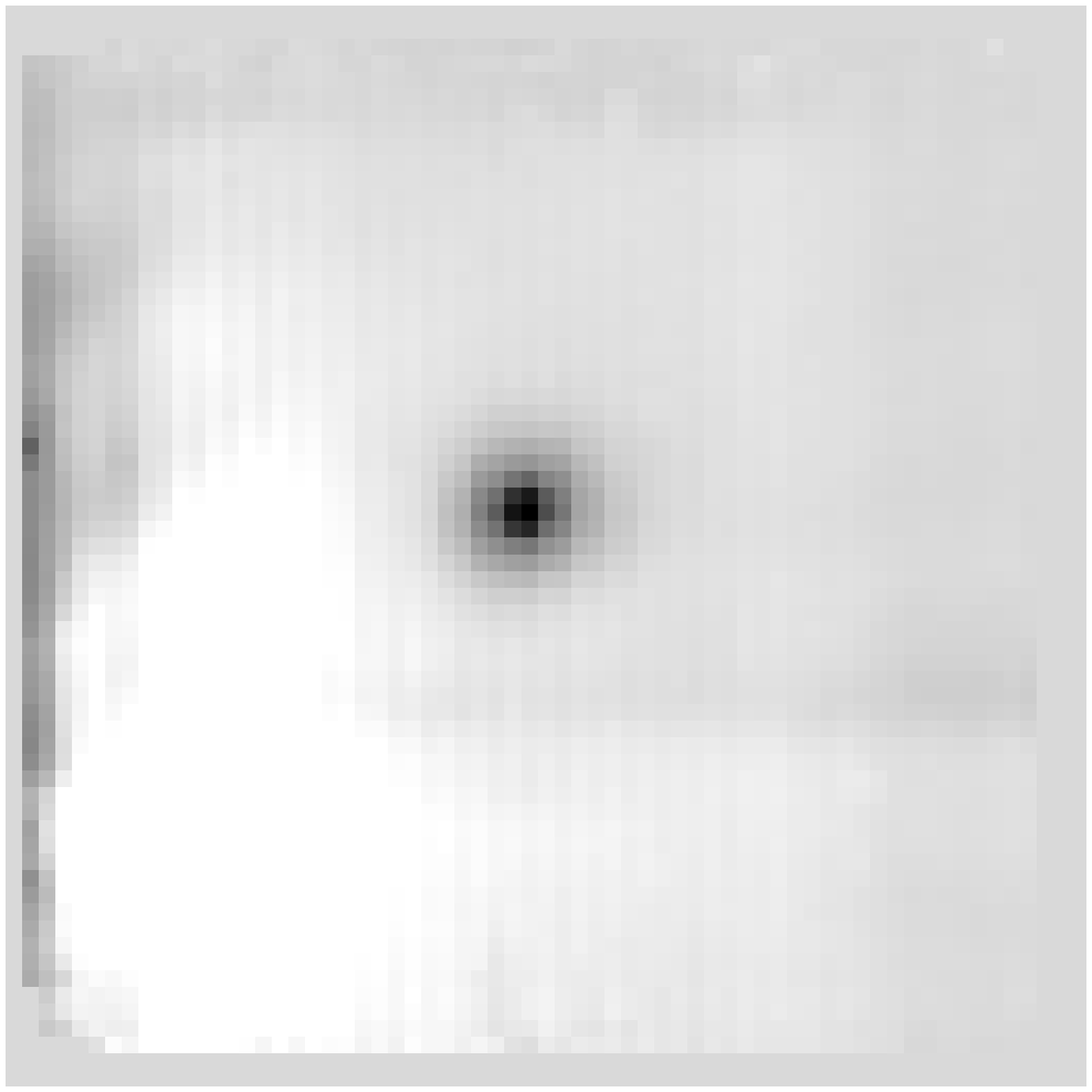}
\includegraphics[bb=80 170 532 622,angle=90,clip]{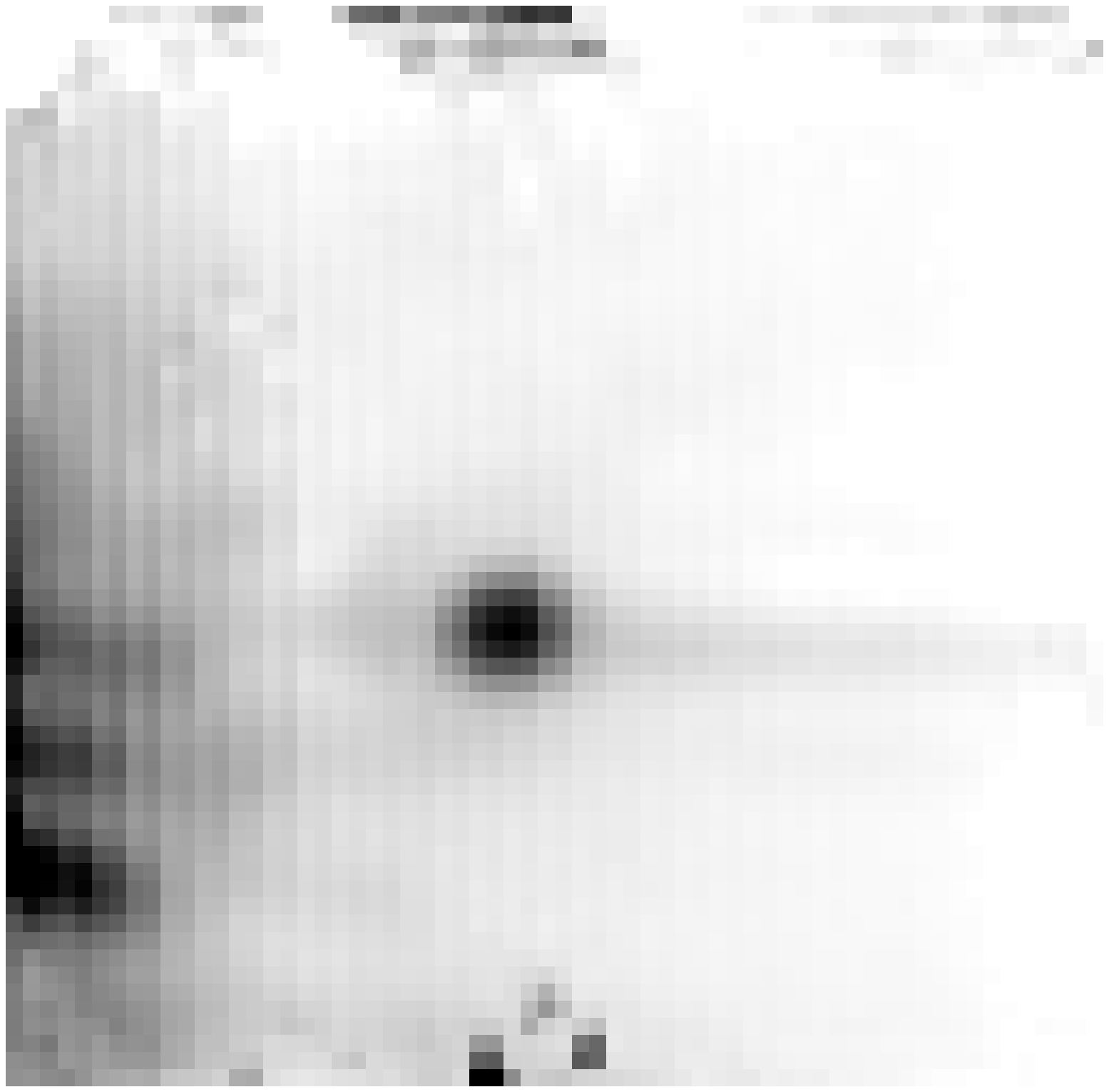}
}
   \caption{\textit{From left to right:} SINFONI J, H and K band cube of 
the GQ Lup companion, integrated over the the wavelength in each cube. \textit{Top row}: before subtraction 
of the halo of GQ Lup A, \textit{ bottom row:} after subtraction. North is up and east 
is left. The FOV is 0.8\arcsec $\times$ 0.8\arcsec}
 \label{images}%
\end{figure}

A mean along the wavelength axis of such a cube is an equivalent of a broad-band image in J, H, or K 
band respectively. In Fig.~\ref{images} we show such images for all three bands. 
The companion is well visible in the centre of each image. 
Depending on the different strehl ratios in each band, the target is still 
contaminated with light from the halo of GQ Lup A, whose core is outside the FOV. This contamination 
is spatially variable since the Airy pattern of the PSF of the primary is wavelength depended, even within 
a given band. This effect is strongest in the K band. 

For a proper extraction of the target spectrum we have to eliminate this contamination as far as possible.
We decided to use the {\sl Starfinder} package of IDL \citep{starfinder} for an empirical PSF fitting of 
the companion in each of the $\sim$2000 images of the three observed bands. For this purpose a template 
PSF is created from the observation of a standard star right before or after the target observation. 
The PSF of the standard star is thereby spatially supersampled from the $\sim$2000 individual images 
of the source. The {\sl Starfinder} algorithm determines the flux of 
the companion in each wavelength bin by 
fitting the previously created template PSF to the target. We found that one PSF template for each band 
is sufficient to ensure high correlation values (usually $\gtrsim$ 0.9) in the fitting process, 
despite the fact that the PSF shape is variable over the covered band, as explained before. 
In an iterative process the smooth background from the PSF halo of GQ Lup A is subtracted as well. 
The lower column in Fig.~\ref{images} shows the images of the companion after this subtraction. 
The remaining speckles in the halo of GQ Lup A can not be filtered by this process. Since we fit a 
well sampled PSF to the target, the contribution from these localized remnants is however neglectable. 

Eventually we obtained a spectrum of the GQ Lup companion from each nodding cycle in each of the three covered bands. 
These spectra are corrected for telluric absorption by the division through the spectrum of an early 
type telluric standard star. The following standard stars where chosen at the observatory to match the 
airmass of the science observations: HIP087140 (B9V) for the J band, HIP082652 (B3III) for the H band, 
and HIP93193 (B9V) for the K band. These stars are intrinsically 
featureless, apart from weak helium and strong hydrogen lines. These lines have been manually fitted and
removed. The resulting spectrum was then multiplied by a blackbody spectrum to correct for the 
continuum slope of each standard star. The effective temperature of the blackbody was retrieved from the 
literature to $T_{eff}$=10500K for the stars of spectral type B9V and $T_{eff}$=16500K for the B3III type
star, respectively. Note, that we are in the Rayleigh-Jeans regime of the standard stars SED. Hence, the
steepness of the continuum slope is not particularly sensitive to small errors in the effective temperature.
After this procedure the resulting spectra of the companion are essentially free of 
telluric absorption lines and corrected for the throughput of the spectrograph.

The individual spectra from each nodding cycle are combined by a weighted mean. The weights are derived 
from the correlation factor of the PSF fitting. The three combined J, H and K band spectra are displayed 
in Fig.~\ref{spectra}.  The Nyquist sampled spectral resolution is R$\sim$2500 in the J-band and R$\sim$4000 
in H and K band. Note that the signal-to-noise ratio of the spectra is $\sim$ 100 in the J band but only 
$\sim$30 for in the H and K band. These values have been retrieved from the standard deviation 
in each spectral bin of the five (J and H band) and eight (K band) individual spectra per band which 
have been reduced separately.
The rather low signal-to-noise values origin from the spectral undersampling (1.5 pix per resolution element). 
Sub-pixel wavelength shifts between the individual nodding cycles induce artificial noise. However, most 
of the small-scale features seen in the spectra are not noise but unresolved absorption lines. 

\section{Spectral synthesis}

We compared our observed spectra with the theoretical template spectra from the GAIA-Cond model v2.0 
\citep{peter05}, updated from \citet{allard01}, with improved molecular dissociation constants, 
more dust species with opacities, spherical symmetry, and a mixing length parameter $2.0 \cdot $H$_{\rm p}$. 
The grid spans a range of  1000...3000~K in effective temperature and -0.5...6.0 dex in surface gravity. 
The metallicity is fixed at [Fe/H]=0.0. The stepsize of the grid is 100K in effective temperature 
and 0.5 dex in surface gravity. Where appropriate we linearly interpolated the spectral grid to refine
the fits to the measured spectra. At around 2600-2700K (the final effective temperature we derived for the companion
to GQ Lup -- see below), the effect of dust on the spectra is still small 
(it starts to affect the spectra massively around 2400K). Furthermore, dusty models of this class were not 
available , thus we feel justified to use the cond models. 
\begin{figure}
\resizebox{\hsize}{!}{
 \includegraphics[bb=10 5 280 195,clip]{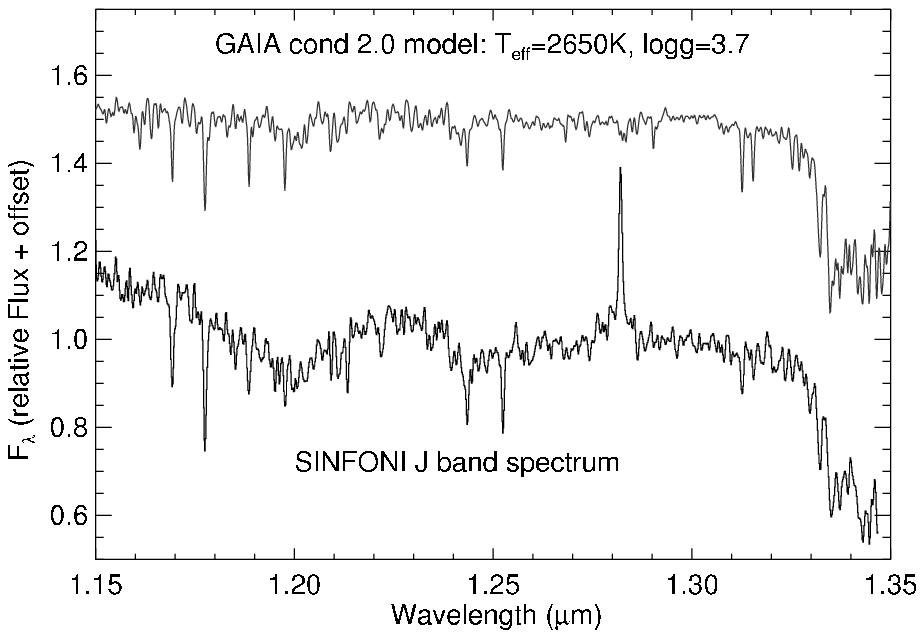}}
\resizebox{\hsize}{!}{
 \includegraphics[bb=10 5 280 195,clip]{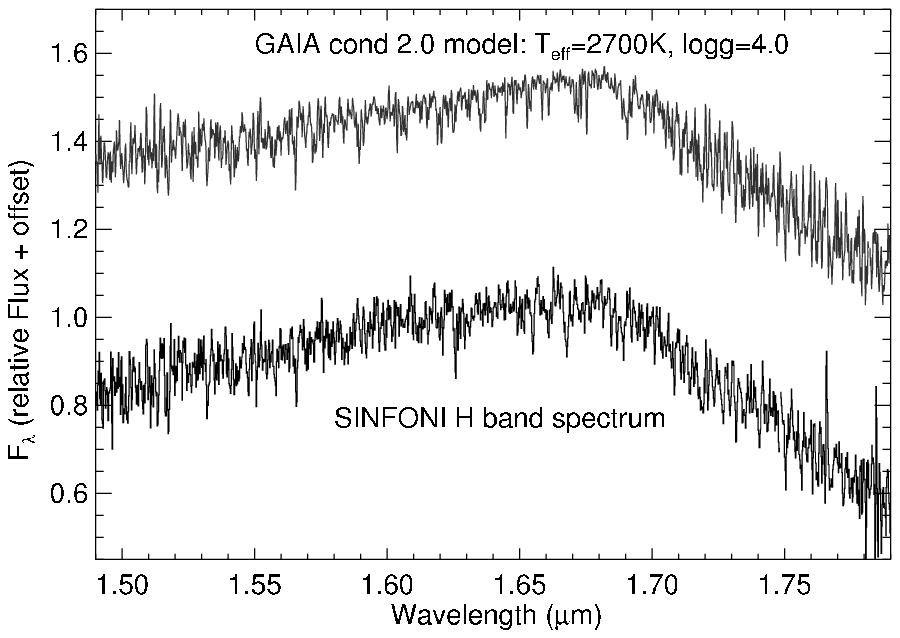}}
\resizebox{\hsize}{!}{
 \includegraphics[bb=10 5 280 195,clip]{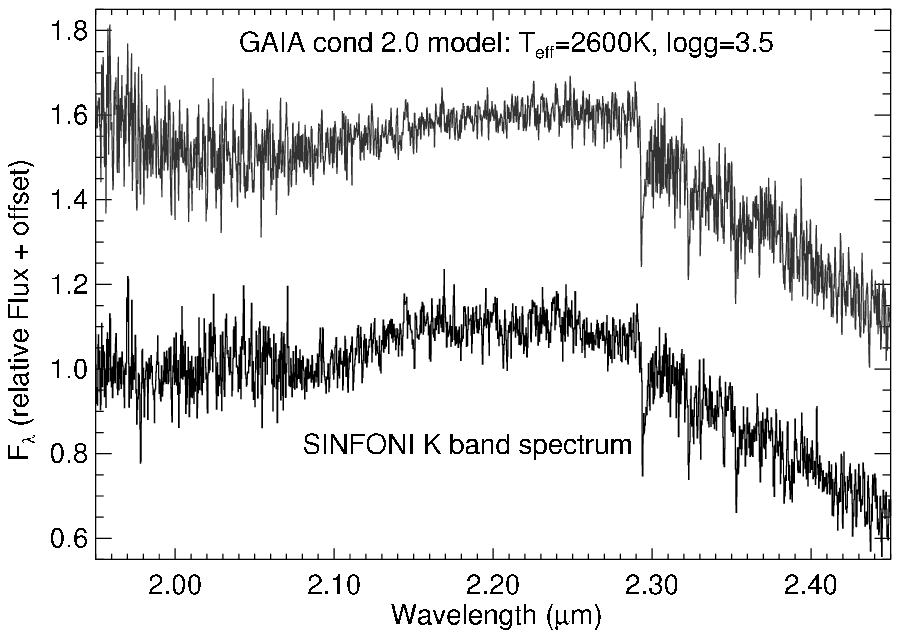}}
   \caption{\textit{From top to bottom:} SINFONI J, H and K band spectra of the GQ Lup companion. 
The best fitting synthetic spectra are overplotted with a relative 
offset of +0.5. The Nyquist sampled spectral resolution is R$\sim$2500 in the J-band 
and R$\sim$4000 in H and K band. Note the strong Pa-$\beta$ emission line in the J-band. }
 \label{spectra}%
\end{figure}
The synthetic spectra are computed for a resolution of 200000. We convolved the spectra with a Gaussian kernel
to reduce the resolution to the one of the SINFONI spectra. Finally the synthetic spectra 
where re-binned to match the spectral sampling of the measured spectra. For the entire grid of synthetic spectra a $\chi^2$-algorithm was applied to determine the best-fitting model spectrum, see Fig.~\ref{spectra}.

\subsection{Spectral features}
\subsubsection*{J band}
A water vapor band absorption longwards of 1.30~$\mathrm{\mu m}$ is 
very well fitted by the models and gives strong constraint on the effective temperature. 
The strong metal lines (K I doublets at 1.169 \& 1.178~$\mathrm{\mu m}$ and at 
1.244 \& 1.256~$\mathrm{\mu m}$ as well as Al I features at 1.313 \& 1.315~$\mathrm{\mu m}$) 
are very sensitive on the effective temperature and the surface gravity. However, their relative 
linestrengths can not be modeled accurately since the model grid has a fixed metallicity at 
[Fe/H]=0.0. Hence these features have been left out of the fitting process. 
The most prominent feature is the Pa $\beta$ emission line at 1.282~$\mathrm{\mu m}$.
This line shows an inverse P-Cygni profile in GQ Lup A. We identify the emission as being
most likely caused by accretion in both objects.

The continuum of the J band shortwards of 1.25$\mathrm{\mu m}$ looks exceptionally bumpy and
is only marginally fitted by the models. The steep decrease in flux from 1.15 to 1.2 $\mathrm{\mu m}$ 
is also seen in the spectra of the young field L dwarf, \object{2MASS J0141-4633}, and most likely due to
VO and FeH absorption \citep{kirk06}. However, in our spectrum it falls in a region that is most 
subject to remaining contamination from GQ Lup A. We thus decided to restrict the fitting process 
to wavelength longer than 1.2 $\mathrm{\mu m}$. Moreover, the increasing slope from 1.20 to 
1.23~$\mathrm{\mu m}$ observed in the GQ Lup companion was also observed in 2MASS J0141-4633 and other young, 
low-gravity objects like \object{KPNO-Tau 4} and \object{G 196-3 B} \citep{Mcgovern} and is due to FeH bands. 
It is also not yet reproduced by our GAIA-cond model and has been consequently disgarded from
the fit.
\subsubsection*{H band}
The continuum slope of the H band exhibits a strong triangular shape that is usually 
identified with low-gravity objects. The nature of this feature is supposedly the H$_2$ collision induced
absorption (CIA) as discussed by \citet[see references therein]{kirk06}. 
The H band is essentially free of strong absoption lines from molecules and metals. 
The constraints on the effective temperature and surface gravity is the lowest among the three bands.
\subsubsection*{K band}
The K band spectrum shows an excellent agreement between model and 
measurement, both, in the slope of the pseudo-continuum and the depth of the most prominent 
spectral features. Again, water vapor absorption at both edges of the spectrum is a good
indicator for the effective temperature. The $^{12}$CO band heads longward of 
2.29~$\mathrm{\mu m}$ are also very well fitted. 
We identify a weak Na I doublet at 2.201~$\mathrm{\mu m}$ which is however slightly better 
fitted by a $\log{g}$ determined from J and H band. We note a non-detection of 
Br $\gamma$ line at 2.166~$\mathrm{\mu m}$. This line is in emission in the spectra of GQ Lup A
where it shows a slightly asymmetric profile.

\subsection{Determination of $T_\mathrm{eff}$ and $\log{g}$}
\begin{figure}
\begin{center}
  \resizebox{0.95\hsize}{!}{\includegraphics[bb= 0 5 283 180,clip]{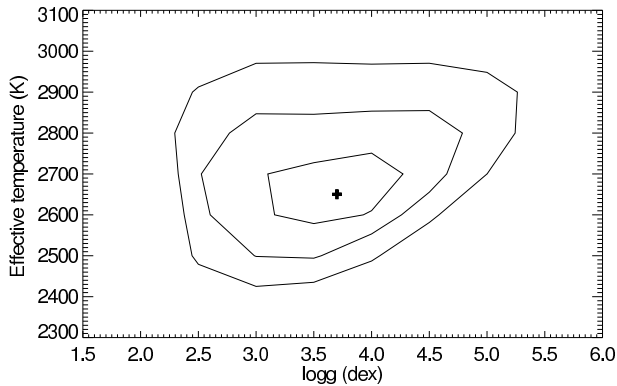}}
\end{center}
   \caption{Results from the fit of GAIA cond synthetic model spectra to the 
SINFONI J, H and K band spectra of the GQ Lup companion. Best value marked with a cross. Contours relate to 1, 2 and 3$\sigma$ errors.}
 \label{contour}%
\end{figure}

The results of the fitting process are displayed in Fig.~\ref{contour} for the combined $\chi^2$
minimisation in a $\log{g}$--$T_\mathrm{eff}$ plane. The derived uncertainties 
(displayed in Fig.~\ref{contour} as contours of 1, 2 and 3$\sigma$) stand for the systematic errors.
We used a Monte-Carlo simulation technique to infer the statistical errors on the determination 
of the model parameters. The uncertainties from the noise in the spectra where found to be much lower
than the uncertainties from fitting errors. This demonstrates that the slope of the continuum is 
the major contraint as long as the spectral lines are not well resolved. 

While the best fit for the J band alone is $T_\mathrm{eff} = 2650~\mathrm{K}$ and 
$\log{g} = 3.70~\mathrm{dex}$, the best fit for the H and K band spectrum have 
slightly different effective temperatures and surface gravities, as can be seen in 
Fig.~\ref{spectra}. Since the J band spectrum has the steepest $\chi^2$ curve, it is 
dominating the final solution, which we determine to be $T_\mathrm{eff} = 2650 \pm 
100~\mathrm{K}$ and $\log{g} = 3.70 \pm 0.5~\mathrm{dex}$.

\section{Results and discussion}

To determine the mass of the GQ Lup companion independently from evolutionary models, 
we need its surface gravity and radius. 
The latter can be derived from a previously determined 
K band magnitude $K_{s} = 13.1 \pm 0.2$~mag \citep{neuh05} and a bolometric correction. 
Based on our new results for the effective temperature, we have to slightly revise the bolometric 
correction used in \citet{neuh05}. From table 6 of \citet{goli} we determine $BC_\mathrm{K} = 
3.00 \pm 0.07$ mag, interpolating over the objects with comparable effective temperature.

With these values, we derive a luminosity of $\log (L/L_{\odot}) = -2.25 \pm 0.24$, slightly 
higher than in \citet{neuh05}, but consistent within the error margin.
From luminosity and temperature, we can then estimate the radius of the companion to be $R = 3.50^{+1.50}_{-1.03}~\mathrm{R_{Jup}}$. Gravity and radius then 
determine the mass, for which we get $\sim 25~\mathrm{M_{Jup}}$, however, 
with large uncertainty. The mass can be as low as $\sim 4~\mathrm{M_{Jup}}$ 
for lower gravity or even as high as $\sim 155 \mathrm{M_{Jup}}$ 
given the large uncertainty in the surface gravity. 

At face value, the mass does not seem to be better contrained now, 
compared with \citet{neuh05}, who gave 1 to 42 Jupiter masses. 
However, we think that due to the low resolution, low signal-to-noise, 
and limited spectral coverage the surface gravity and effective temperature 
may have been underestimated in \citet{neuh05}.
Even though our mass determination is independent from evolutionary models, 
it still depends on the atmospheric models. Until the GAIA model is tested 
and confirmed in the very low-mass and young age regime, our results are to 
be seen uncertain.
However, we argue that the accuracy of a mass determined from spectral synthesis 
should be higher than from evolutionary models since the latter should use synthetic 
atmospheric models itself as an input parameter.


Stassun et al. (2006) found an eclipsing double-lined spectroscopic binary, 
2MASS J05352184-0546085 (hereafter 2M0535), consisting of two brown 
dwarfs in the Orion star forming region. Their masses could be determined dynamically, 
hence independently from evolutionary models and model atmospheres. 
This makes 2M0535 most suitable for a direct comparison with the GQ Lup companion.

Note, that the more massive object in 2M0535 is, however, cooler, which is not 
expected from standard models (e.g. Lyon or Tucson groups). Stassun et al. (2006) 
also note a systematic uncertainty of $\pm 300$ K in the absolute temperatures, 
and can give only a precise temperature difference or fraction.

We note that the companion to GQ Lup has roughly the same age but is fainter and smaller 
than both components in 2M0535 \citep{stassun}, see Table~1. The effective temperature is 
comparable with the cooler component in this double. Judging from the temperature and 
radius of the nearly coeval objects, the GQ Lup companion must be lower in mass than 
each of the components in 2M0535, i.e. $\le 36~~\mathrm{M_{Jup}}$.

\begin{table}[b!]
\caption{Comparing the GQ Lup companion to 2M0535 A and B}
\label{tab1}
\begin{tabular}{lccc}
\hline\hline
Parameter                  & \object{GQ Lup b}/B (a)      & 2M0535 A           & 2M0535 B \\ \hline
$T_{eff}$ [K]              & $2650 \pm 100$    & $2650 \pm 100$ (b) & $2790 \pm 105$ (b) \\
$\log~(L_{bol}/L_{\odot})$ & $-2.25 \pm 0.24$  & $-1.70 \pm 0.11$   & $-1.74 \pm 0.11$   \\
Age [Myr]                  & 0 -- 2            & 0 -- 3             & 0 -- 3             \\
Radius [R$_{\odot}$]             & 0.36$^{+0.15}_{-0.11}$ & $0.669 \pm 0.034$  & $0.511 \pm 0.026$  \\ 
Mass [M$_{\odot}$]               & 0.025$^{+0.120}_{-0.021}$ (c) & $0.054 \pm 0.005$  & $0.034 \pm 0.003$  \\ \hline
\end{tabular}
Remarks: (a) GQ Lup b, if planet or planet candidate; GQ Lup B, if brown dwarf or star. \\
(b) $\pm 300$ K systematic uncertainty (Stassun et al. 2006). \\
(c) Regarding the upper limit: our companion is fainter and smaller than both 2M0535 A and B, 
hence lower in mass than the components of 2M0535.
\end{table}

Finally we note two recent results on GQ Lup:

Marois et al. (2006) fit the RIJHKL-band spectral energy distribution of 
the GQ Lup companion with a GAIA model to obtain physical parameters. 
They find a radius of $0.38 \pm 0.05~\mathrm{R_{\sun}}$, and an effective 
temperature of $2335 \pm 100$~K. Hence, the authors confirm temperature and 
luminosity of \citet{neuh05}. Their modelfits are reported to be gravity insensitiv 
and a $\log{g}$ of 3-4 is assumed. They use the evolutionary models from the Lyon 
group to obtain a mass of 10 to 20 Jupitermasses.

McElwain et al. (2006) obtained an integral field J and H-band spectrum with 
OSIRIS at Keck, which is lower in dynamic range, resolution, S/N, and spectral coverage 
than ours. They confirm the luminosity, and temperature from \citet{neuh06} and 
obtain a mass of 10 to $40~\mathrm{M_{Jup}}$.

Both authors use the models of the Lyon group. We argue that these models
are not applicable for objects up to $\sim 10$ Myrs, following the discussion
in \citet{chab}.

Based on our and otherwise published results, one still cannot decide, whether
the companion to GQ Lup is a massive planet or a low-mass brown dwarf.

\begin{acknowledgements}

We would like to thank the staff of the VLT at Paranal for the execution of the 
observations. We also thank Eike W. Guenther for useful discussions about the layout 
of the measurement. Finally we thank the anonymous referee for the thorough review 
and the helpful comments.
 
\end{acknowledgements}

\end{document}